\documentclass[twocolumn,showpacs,preprintnumbers,amsmath,amssymb,pre]{revtex4}

\usepackage{graphicx}
\usepackage{dcolumn}
\usepackage{bm}
\usepackage{enumerate}

\begin{document}

\title{Time evolution of the reaction front in a subdiffusive system}

\author{Tadeusz Koszto{\l}owicz}
 \email{tkoszt@pu.kielce.pl}
 \affiliation{Institute of Physics, \'Swi\c{e}tokrzyska Academy,\\
         ul. \'Swi\c{e}tokrzyska 15, 25-406 Kielce, Poland.}

\author{Katarzyna D. Lewandowska}
 \email{kale@amg.gd.pl}
 \affiliation{Department of Physics and Biophysics, Medical University of
         Gda\'nsk,\\ ul. D\c{e}binki 1, 80-211 Gda\'nsk, Poland.}

\date{\today}

\begin{abstract}
Using the quasistatic approximation we show that in a subdiffusion--reaction system with arbitrary non--zero values of subdiffusion coefficients, the reaction front $x_{f}(t)$ evolves in time according to the formula $x_{f}(t)= K t^{\alpha/2}$, with $\alpha$ being the subdiffusion parameter and $K$ which is controlled by the subdiffusion coefficients. The parameter $K$ is determined by the equation derived in this paper. To check correctness of our analysis, we compare analytical functions derived in this paper with the results obtained numerically for the subdiffusion-reaction equations.
\end{abstract}

                            
\maketitle

\section{Introduction}

The diffusion--reaction system with two initially separated diffusing particles of spices $A$ and $B$ reacting according to the formula $m'A+n'B\rightarrow P({\rm inert})$ has been intensively studied during past years \cite{11,12,17,13,14,15,ckd,ara,18,19,cd,22,24}. As the diffusion-reaction equations describing the system are nonlinear, it is difficult to solve them and their general solutions remain unknown (except of very special cases). Thus, to simplify the calculations one usually uses various approximations, such as the quasistationary approximation \cite{11,12,17}, the scaling method \cite{11,13,14,15,ckd}, or the perturbation one \cite{18,19}. Using these methods, there were derived characteristic functions of the system which include: the time evolution of the reaction front $x_f(t)$, the width of the reaction zone $W_{\rm R}(t)$ or the width of the depletion zone $W_{\rm Dep}(t)$ \cite{11,12,13,14,15} which all appear to be the power functions of time $f(t)=K t^\gamma$. The results were confirmed by numerical calculations and simulations \cite{14,15,17}. However, as the methods of extracting the power functions are not based on analytical solutions of subdiffusion-reaction equations (not even on their approximately forms) the proportionality coefficient $K$ is unknown. The coefficient carries dynamical information about the system e.g. how the diffusion coefficient influences the process. As far as we know, there were only a few attempts to determine of $K$ by means of the quasistationary approximation \cite{12,17}. 

The situation is more complicated in the case of the subfiffusion system since the equations describing the system contain a derivative of fractional order. 
Subdiffusion occurs in systems where mobility of particles is
significantly hindered due to internal structure of the medium, as in porous media or gels \cite{mk,kdm}. The subdiffusion is characterized by a time dependence of the mean square displacement of transported particle $\left\langle \Delta x^2\right\rangle=2D_\alpha t^{\alpha}/\Gamma\left(1+\alpha\right)$, where $D_\alpha$ is the subdiffusion coefficient measured in the units $m^2/s^{\alpha}$ and $\alpha$ is the subdiffusion parameter which obeys $0<\alpha<1$. For $\alpha=1$ one deals with the normal diffusion. 
Since no explicit solutions of the nonlinear (sub)diffusion-reaction equations are known, one commonly considers a simplified system, for example a one in which diffusion coefficients of both reactants are assumed to be equal to each other. There is assumed that the some characteristic functions are the same in the system with any simplified assumptions \cite{27a,10}. 

The problem is to choose a method to study the subdiffusion-reaction equations. The scaling method dose not allow one to determine $K$ unless special conditions are taken into account. The perturbation method is of small efficiency because the first order correction is often insufficient, while the higher order corrections are hard to obtain even in the case of normal diffusion. So, there are a few problems solved using this method \cite{18,19}. An alternative method is the quasistationary one. In the case of normal diffusion-reaction system it is based on the assumption that process proceeds so slowly that changes of concentration of transported substance are small in some regions \cite{12,17}. Since the subdiffusion process is significantly slower than the normal diffusion one, we expect that the quasistationary approximation is also applicable to the subdiffusive case. Thus, we adopt the method in this study. The scaling method and the quasistationary approximation one are often treated as equivalent to each other. We note however that the equivalence holds only in the long time limit \cite{10}. At shorter times applicability of the quasistationary method does not imply applicability of scaling one and vice versa (this problem will be discussed in \cite{kosztlew}).

In this paper we find that the time evolution of the reaction front is given by the formula $x_{f}(t)=K t^{\alpha/2}$
for a system with arbitrary non--zero values of the subdiffusion coefficients. The coefficient $K$ fulfills the special equation derived in this paper. Our analytical results are confirmed by the numerical solutions of subdiffusion--reaction equations.

\section{\label{sys}The system}

A real system is usually three--dimensional, but we assume that it
is homogeneous in the plane perpendicular to the $x$ axis.
Therefore, we involve only one space variable $x$ into considerations. The subdiffusion--reaction equations are 
		 \begin{equation}\label{eq2}
\frac{\partial }{\partial t}C_{i}(x,t)=D_{i}\frac{\partial^{1-\alpha}}{\partial t^{1-\alpha}}\frac{
\partial^{2}}{\partial x^{2}}C_{i}(x,t)-d_iR_{\alpha}(x,t),
    \end{equation}
where $i=A,B$, $C_{i}$ denotes the concentration of the diffusing
particles of species $i$, $D_{i}$ -- the subdiffusion
coefficient, $d_A=m$, $d_B=n$, the parameters $m$ and $n$ occur in the reaction term (Eq.~(\ref{eq4}) below); the Riemann--Liouville fractional time derivative is defined for the
case of $0<\alpha<1$ as
    \begin{displaymath}
\frac{d^\alpha f(t)}{dt^\alpha}=\frac{1}{\Gamma(1-\alpha)}\frac{d}{d t}\int_{0}^{t}d\tau\frac{f(\tau)}{(t-\tau)^\alpha}.
    \end{displaymath}
Throughout this paper we assume that both of the reactants are mobile $D_{A},D_{B}>0$. Let us note that the choice of the reaction term is not obvious \cite{9,27a,10,sbsl,s,hw,10a}. The reaction term, which we involve into considerations and which was used to study the subdiffusion--reaction system in \cite{27a,10}, is
    \begin{equation}\label{eq3}
R_{\alpha}(x,t)=\frac{\partial^{1-\alpha}}{\partial
t^{1-\alpha}}R(x,t) ,
    \end{equation}
where the term $R(x,t)$ within the mean field approximation reads
    \begin{equation}\label{eq4}
R(x,t)=kC_A^{m}(x,t)C_B^{n}(x,t),
    \end{equation}
$k$ is the reaction rate and the parameters $m$ and $n$ are determined experimentally. 

\begin{figure}[h!]
\centering
\includegraphics[scale=0.95]{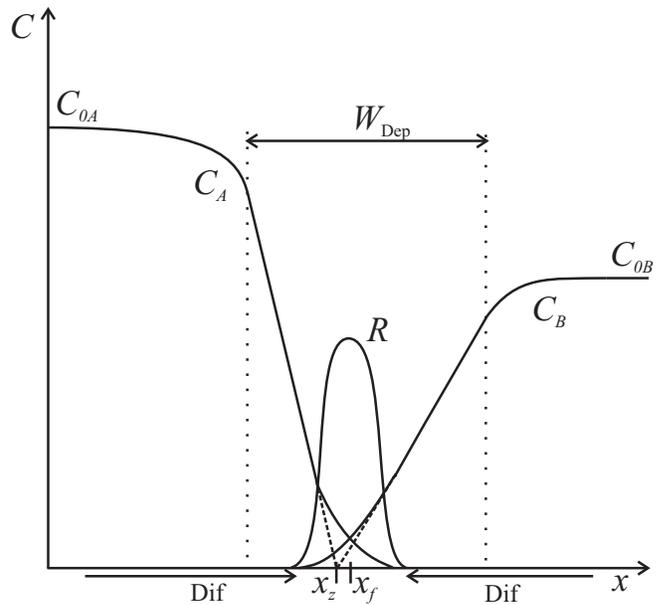}
\caption{\label{Fig.1}Schematic view of the system under
considerations; $x_{f}(t)$ is the reaction front, ${\rm `Dep'}$ and
${\rm `Dif'}$ denote the depletion zone and the
diffusion region, respectively.}
\end{figure}

We assume that the particles of reactants $A$ and $B$ are initially separated from
each other. Thus, the initial conditions are
\begin{eqnarray}
    C_A(x,0)&=&\left\{ \begin{array}{cc}
             C_{0A}, & x<0 \\
             0, & x>0
           \end{array} \right.,\label{eq9a}\\
    C_B(x,0)&=&\left\{ \begin{array}{cc}
             0, & x<0 \\
             C_{0B}, & x>0
           \end{array} \right. .\label{eq9b}
\end{eqnarray}           
It was observed \cite{11,12,17,13,14,15} that when
the process starts, there appear characteristic regions (see
Fig.~\ref{Fig.1}): the depletion zone ${\rm `Dep'}$, which is
defined as a region where the concentrations are significantly
smaller than the initial ones ($C_A\ll C_{0A}$ and $C_B\ll
C_{0B}$), the reaction region where the production of
particles $P$ is significant ($R(x,t)>0$), and the
diffusion region ${\rm `Dif'}$,
where the reaction term $R(x,t)$ is close
to zero and the particle transport appears to be almost subdiffusive
(i.e. without chemical reactions).

For the normal diffusion the widths of the depletion zone $W_{{\rm
Dep}}$ and the reaction region $W_{{\rm R}}$ grow as the power
functions of time \cite{11,12,17,13,14,15,ckd}, $W_{{\rm Dep}}\sim t^{\theta}$, with $\theta=1/2$, and $W_{{\rm R}}\sim t^{\mu}$, where $\mu<\theta$. The value of parameter $\mu$ depends on the system under study.
For the system where the reactants $A$ and $B$ are mobile, there is
$\mu=1/6$ and for the system with a mobile reactant $A$ and a
static reactant $B$ we have $\mu=(m-1)/2(m+1)$, where $m$ is the parameter occurring in the
reaction term $R$ in Eq.~(\ref{eq4}) \cite{13} (see also
\cite{ckd,cd}). As was reported in \cite{ara,10}, $W_{\rm R}$ evolve in time according to the power functions also for the subdiffusive--reaction system with $\mu =\alpha/6$.

An important characteristics of the system under consideration is the time evolution of the reaction front $x_{f}(t)$. It is defined as a point where the reaction term
$R(x,t)$ reaches its maximum $R(x_{f}(t),t)=max$ or, as argued in \cite{14}, for $R\sim C_AC_B$ it is defined by the relation $C_A(x_f(t),t)=C_B(x_f(t),t)$ and in more general situation by $C_A(x_f(t),t)/m=C_B(x_f(t),t)/n$ \cite{ckd}. Unfortunately, these definitions are difficult to apply for the numerically obtained concentrations. In the following, we use the definition of the reaction front as
	\begin{equation}\label{defxf}
x_f(t)=\frac{\int x R(x,t)dx}{\int R(x,t)dx} .
	\end{equation}
Although the relations defining the reaction front are not always equivalent to each other, all of them provide to $x_f$ lying inside the reaction region, and in the long time limit the definitions lead to the power functions of time.

For the normal diffusion there is the dependence \cite{11,12,17,13,14,15}
    \begin{equation}\label{xfg}
x_{f}(t)\sim t^{\gamma},
    \end{equation}
with $\gamma=1/2$. It was shown in \cite{10} by
means of the scaling method that the relation (\ref{xfg}) with $\gamma=\alpha/2$ holds for
the subdiffusive system where the subdiffusion coefficients of
reactants are equal to each other.

\section{Quasistatic approximation}

The quasistatic approximation assumes that the concentration profile is
a slowly varying function of time in a given region. Thus, 
the time derivation is small and consequently, the r.h.s. of (sub)diffusion 
equation (\ref{eq2}) is also small in the region. It requires
	\begin{equation}\label{r2}
\frac{\partial^{1-\alpha}}{\partial t^{1-\alpha}}\frac{\partial^2}{\partial x^2}C_{A,B}(x,t)\approx R_\alpha(x,t).
	\end{equation}

Since the reaction term is relatively large in the reaction zone,
the quasistatic approximation holds in this zone under the condition  
	\begin{equation}\label{r3}
D\frac{\partial^{1-\alpha}}{\partial t^{1-\alpha}}
\frac{\partial^{2}}{\partial x^{2}}C_{A,B}(x,t)\gg\frac{\partial }{\partial t}C_{A,B}(x,t).
	\end{equation}
We note that the condition (\ref{r3}) is fulfilled when the concentration profiles are given in the scaling form \cite{10}. 

In the diffusive region where $R_\alpha(x,t)\approx 0$, the quasistatic approximation
is applicable when the concentration is a linear function of $x$, as the r.h.s. of
Eq. (\ref{eq2}) then vanishes. The regions outside the reaction zone, where the
concentration linearly varies with $x$, determine the borders of the quasistatic
region. The solution of subdiffusion equation without chemical reactions works here. 

In the studies of the normal diffusion with reactions, one introduces the quasistatic approximation referring to the equilibration time $\tau_{\rm F}$ \cite{12,ckd,cd}. The parameter $\tau_{\rm F}$ is of order of the average time which is needed for the substance to spread over the interval of length $W_{\rm R}$ when the substance flows from outside of the interval.
For the normal diffusion-reaction system this parameter was estimated from the relation $\left\langle\Delta x^2\right\rangle\sim t$. Taking $\left\langle\Delta x^2\right\rangle\sim W^2_{\rm R}$ and $t\sim \tau_F$ one gets $\tau_{\rm F}\sim W_{\rm R}^2$. For the subdiffusive system the relation $\left\langle\Delta x^2\right\rangle\sim t^\alpha$ provides 
	\begin{equation}\label{tf}
\tau_{\rm F}\sim W_{\rm R}^{2/\alpha}.
	\end{equation}

As for the normal diffusion case, let us assume that the relative change of the flux $J$ fulfills the relation $dJ/J=dt/\tau_{\rm J}$  which gives
	\begin{equation}\label{tj}
(\tau_{\rm J})^{-1}\sim \frac{d({\rm log}J)}{dt}.
	\end{equation} 
The balance between the subdiffusion term and the reaction one is achieved when the equilibration time $\tau_{\rm F}$ of the reaction region is negligibly small comparing to the time ($\tau_{\rm J}$) of relative change of the flux  in the long time limit. So, the quasistatic approximation is applicable when 
	\begin{equation}\label{r4}
\frac{\tau_{\rm F}}{\tau_{\rm J}}\rightarrow_{t\rightarrow\infty}0.
	\end{equation}
The quasistatic region is usually defined as a region where at least one of the conditions (\ref{r2}), (\ref{r3}) or (\ref{r4}) is fulfilled. As far as we know, the equivalence of these definitions have not been proven yet. In our considerations we use the relation (\ref{r2}) as the definition of quasistatic approximation and we further show that the conditions (\ref{r4}) is fulfilled when Eq.~(\ref{r2}) is assumed.

\section{Time evolution of $W_{{\rm R}}$ and $W_{{\rm Dep}}$}\label{wrwd}

As in the normal diffusion reaction system, to find the widths of appropriate region we assume that the parameters $p=D_B/D_A$ and $q=C_{0B}/C_{0A}$ are the irrelevant parameters of the system. This means that the results obtained for $p=1$ and/or $q=1$ are qualitatively equivalent to the one with $p\neq 1$ and/or $q\neq 1$, except of very few obvious cases (for example, when $p=q=1$ the reaction front does not move). In this section we derive the time evolution of the widths of the reaction region $W_{\rm R}$ and the depletion zone $W_{{\rm Dep}}$ under condition $p=q=1$.

At first we argue that the assumption $\mu<\theta$ is correct not
only for the diffusive but for the subdiffusive systems as well. 
In \cite{10} there was found
that $\theta = \alpha/2$ and $\mu = \alpha/6$ by means of the simplified scaling method. We confirm the above relation, using the method
already applied to the normal diffusion--reaction system
\cite{24}.

\begin{figure}[h!]
\centering
\includegraphics[scale=0.95]{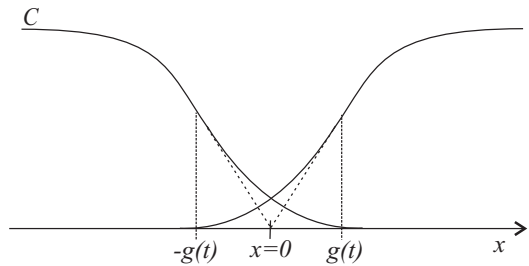}
\caption{The symmetrical system with static reaction front
$x_{f}(t)=0$. The continuous lines denote the concentrations $C$ for
the system under considerations, the dashed one -- for the system
with the fully absorbing wall $C_{{\rm abs}}$ located at $x=0$.}\label{Fig.2}
\end{figure}

Let us assume that the initial concentrations and subdiffusion
coefficients of both reactants are equal to each other $C_{0A}=C_{0B}\equiv C_{0}$ and $D_{A}=D_{B}\equiv D$.
Due to the symmetry of the system, the reaction front will not
change its position $x_{f}(t)=0$. Proceeding as for the system with the normal diffusion
\cite{24}, we assume to further simplify of the calculations that
the concentrations of $A$ and $B$ particles can be given as
    \begin{displaymath}
C_A(x,t)=C_{A \rm abs}(x,t)+\delta C_A(x,t),
    \end{displaymath}
    \begin{displaymath}
C_B(x,t)=C_{B \rm abs}(x,t)+\delta C_B(x,t),
    \end{displaymath}
where $C_{A {\rm abs}}$ and $C_{B {\rm abs}}$ are the solutions of the pure subdiffusive equation in the system with the perfectly absorbing wall located at $x=0$ and
$\delta C_A$ and $\delta C_B$ are the corrections (see Fig.~\ref{Fig.2}). Symmetry of the
system ensures that $C_A(x,t)=C_B(-x,t)$, which provides $\delta
C_A(x,t)=\delta C_B(-x,t)$. For a perfectly absorbing wall
placed at $x=0$, the concentration profiles vanish at the wall
$C_{A \rm abs}(0,t)=C_{B \rm abs}(0,t)=0$. After calculations, we obtain
\begin{eqnarray}\label{eq27}
\lefteqn{C_{A \rm abs}(x,t)=}\nonumber\\
& & C_{0}\left[1 -\frac{2}{\alpha}H^{1 0}_{1 1}
  \left(\left(\frac{-x}{\sqrt{Dt^{\alpha}}}\right)^{2/\alpha}
            \left| \begin{array}{cc}
             1 & 1 \\
             0 & 2/\alpha
           \end{array}\right. \right)
           \right],
\end{eqnarray}
for $x<0$, and
\begin{eqnarray}\label{eq28}
\lefteqn{C_{B \rm abs}(x,t)=}\nonumber\\
&&C_{0}\left[1-\frac{2}{\alpha} H^{1 0}_{1 1}
  \left(\left(\frac{x}{\sqrt{Dt^{\alpha}}}\right)^{2/\alpha}
            \left| \begin{array}{cc}
             1 & 1 \\
             0 & 2/\alpha
           \end{array}\right. \right)
           \right],
\end{eqnarray}
for $x>0$, where $H$ denotes the Fox function, which can be
expressed as the series \cite{23}
    \begin{equation}\label{eq29}
 H^{1 0}_{1 1} \left(u \left|
      \begin{array}{cc}
             1 & 1 \\
             p & q
     \end{array}
 \right. \right)= \frac{1}{q}u^{p/q}
  \sum^\infty_{j=0}\frac{(-1)^{j}}{j!\Gamma(1-p/q-j/q)}u^{j/q}.
    \end{equation}

Substituting $C(x,t)\equiv C_i(x,t)$, $\delta C(x,t)\equiv \delta C_i(x,t)$, $C_{\rm abs}(x,t)\equiv C_{i \rm abs}(x,t)$, where $i=A$ for $x<0$ and $i=B$ for $x>0$,
to the subdiffusion--reaction equation and taking into account that
$C_{\rm abs}$ fulfills the subdiffusion equation without
chemical reactions, we get
    \begin{eqnarray*}
\frac{\partial }{\partial t}\delta
C(x,t)&=&\frac{\partial^{1-\alpha}}{\partial
t^{1-\alpha}}\left[D\frac{
\partial^{2}}{\partial x^{2}}\delta C(x,t)\right.\\
& & \left. -k\left[C_{\rm abs}(x,t) +\delta C(x,t) \right]\delta
C(x,t)\right].
    \end{eqnarray*}
The limits of the reaction region occur for $x$ where
$\delta C$ is close to zero. In this region one can neglect the
term $(\delta C)^{2}$ in the above equation. Moreover, in the long
time limit we can approximate the Fox functions present in
Eqs.~(\ref{eq27}) and (\ref{eq28}) by the expression
$C_{\rm abs}(x,t)=a|x|/t^{\alpha/2}$, where
$a=C_{0}k/\Gamma(1-\alpha/2)\sqrt{D}$. So, we get
    \begin{equation}\label{eq32}
\frac{\partial }{\partial t}\delta
C(x,t)=\frac{\partial^{1-\alpha}}{\partial
t^{1-\alpha}}\left[ D\frac{
\partial^{2}}{\partial x^{2}}\delta C(x,t)
-\frac{a|x|}{t^{\frac{\alpha}{2}}}\delta C(x,t)\right].
    \end{equation}
As in the normal diffusion--reaction system \cite{24}, we assume that
    \begin{equation}\label{eq32a}
 \frac{\partial}{\partial t}\delta C(x,t)=0 .
    \end{equation}
From Eqs.~(\ref{eq32}), (\ref{eq32a}) and the relation \cite{20}
	\begin{equation}\label{pochtni}
\frac{d^{\beta}t^{\nu}}{dt^{\beta}}=
\frac{\Gamma(\nu+1)}{\Gamma(\nu+1-\beta)}t^{\nu-\beta},\qquad \nu>-1,
	\end{equation}
we obtain
   \begin{equation}\label{eq34}
D\frac{
\partial^{2}}{\partial x^{2}}\delta C(x,t)-\frac{a|x|}{t^{\frac{\alpha}{2}}}\delta
C(x,t)=\frac{A(x)}{t^{\alpha}},
    \end{equation}
where $A(x)$ is the arbitrary function of $x$ only. In the long time limit the r.h.s. of Eq. (\ref{eq34}) can be neglected. Let us note that it is another justification to equal the r.h.s. of above equation to zero. To determine the function $A(x)$ we observe that $\delta C$ has a significant value in a finite region limited by the
depending on time points $-g(t)$ and $g(t)$, which lie inside of the
depletion zone (see Fig.~\ref{Fig.2}). Thus, we have
\begin{displaymath}\label{eq34a}
\delta C(-g(t),t)\approx\delta C(g(t),t)\approx 0
\end{displaymath}
and
\begin{displaymath}
\left. \frac{\partial^{2}\delta C(x,t)}{\partial
x^{2}}\right|_{x=-g(t)}\approx\left. \frac{\partial^{2}\delta
C(x,t)}{\partial x^{2}}\right|_{x=g(t)}\approx 0 .
\end{displaymath}
Since the left hand side of the Eq. (\ref{eq34}) is close to
zero for $|x|>g(t)$, the additional boundary conditions are
\begin{equation}\label{eq34c}
A(-g(t))=A(g(t))=0 .
\end{equation}
The above equations appear to be the boundary conditions for the
function $A$, which cannot depend on time (in contrary to the
boundary conditions (\ref{eq34c})). Thus, there is the only
solution $A(x)\equiv const\equiv 0$. 

Solving the equation
(\ref{eq34}) with the right side equal to zero, we find
    \begin{displaymath}
\delta C(x,t)=f(t){\rm Ai}\left(\lambda
\frac{x}{t^{\alpha/6}}\right) ,
    \end{displaymath}
where ${\rm Ai}$ denotes the Airy function, which can be
approximated by the following expression for large~$u$
    \begin{displaymath}
{\rm Ai}(u)\simeq
\frac{1}{2\sqrt{\pi}u^{1/4}}\exp\left[-\frac{2u^{3/2}}{3}\right].
    \end{displaymath}
To obtain the function $f(t)$
we assume that it is the power function of time $f(t)\sim
t^{\lambda}$. Putting the function $f$ to Eq.~(\ref{eq32}) and using
(\ref{eq32a}), we obtain $\lambda=-\frac{\alpha}{3}$. Comparing
Eq.~(\ref{eq32}) with (\ref{eq2}) and (\ref{eq3}), we get
    \begin{equation}\label{eq37}
R(x,t)=\frac{a|x|}{t^{\frac{\alpha}{2}}}\delta C(x,t) .
    \end{equation}
Substituting Eq.~(\ref{eq37}) to Eq.~(\ref{eq32}) we obtain
    \begin{equation}\label{eq38}
 R(x,t)\sim
t^{-2\alpha/3}\left(\frac{x}{t^{\alpha/6}}\right)^{3/4}
\exp\left[-\frac{2}{3}\left(\frac{\lambda
x}{t^{\alpha/6}}\right)^{3/2}\right].
    \end{equation}
As the width of the reaction region is defined by the relation \cite{ckd}	
	\begin{equation}\label{defWR}
W_{\rm R}^2(x,t)=\frac{\int (x-x_f(t))^2 R(x,t) dx}{\int R(x,t) dx}	.
	\end{equation}
it is easy to see that substituting (\ref{eq38}) to (\ref{defWR}) with $x_f\equiv0$, we obtain $W_{{\rm R}}\sim t^{\alpha/6}$.

Since the width of the depletion zone is defined by the conditions
$C_{i}\ll C_{0i}$, $i=A,B$, from Eqs. (\ref{eq27}) and
(\ref{eq28}) we get $W_{{\rm Dep}}\sim
t^{\alpha/2}$. Thus, the relation $\mu<\theta$ is fulfilled for
the system where the subdiffusion coefficients of the reactants
are equal to each other. We assume that this relation holds for the system
with any non--zero values of the subdiffusion coefficients.

\section{Concentration profile in ${\rm Dif}$ region}\label{wdif}

Since $W_{\rm R}\ll W_{{\rm Dep}}$, 
the reaction region plays a role of a partially absorbing wall
with respect to the depletion zone. We find the concentration
profiles in the region outside the reaction one as a solution of
the subdiffusion equation in the system with partially absorbing
wall.

We find here the solutions of the subdiffusion equation without
chemical reactions (Eq. (\ref{eq2}) with $R_{\alpha}(x,t)\equiv0$) for the system with partially
absorbing wall. To calculate the concentration profiles, we use the
integral formula
    \begin{equation}\label{eq41}
C(x,t)=\int G(x,t;x_{0})C(x_{0},0)dx_{0},
    \end{equation}
where $G(x,t;x_{0})$ denotes the Green's function for the
subdiffusion equation. From a macroscopic point of
view, the Green's function is interpreted as a concentration
profile of the $N$ particles (divided by $N$) which are
instantaneously produced and start from the position $x_{0}$ at an
initial moment $t=0$. It is also interpreted as a probability
density of finding a particle in a point $x$ at time $t$ under
the condition that the particle is located in the position $x_{0}$
at the initial moment $t=0$.

There is a problem to set the boundary conditions at the partially
absorbing wall. To obtain the Green's function one can use the
method of images. The standard method of images has been applied for
the diffusive system with fully absorbing or fully reflecting
wall \cite{25}. Then, one replaces the wall by a fictitious
instantaneous point source of the particles (IPS) in such a manner
that the concentration profile generated by all IPS behaves as in
the system with the wall. In the system with the fully reflecting wall,
the flux vanishes at the wall. In this case the Green's function can
be obtained by replacing the wall by the auxiliary IPS of the same
strength in the position symmetric to the initial point $x_{0}$ with
respect to the wall
    \begin{equation}\label{eq42a}
G(x,t;x_{0})=G_{0}(x,t;x_{0})+G_{0}(x,t;-x_{0}) ,
    \end{equation}
where $G_{0}$ denotes the Green's function for homogeneous system. In the case of fully absorbing wall the concentration vanishes at the wall. The Green's function is
then a difference of IPS placed at $x_{0}$ and $-x_{0}$, which
gives
    \begin{equation}\label{eq42}
G(x,t;x_{0})=G_{0}(x,t;x_{0})-G_{0}(x,t;-x_{0}).
    \end{equation}
Sometimes the boundary conditions are not given explicitly by an
equation, but they are postulated in a heuristic form. In such a
case there is a possibility to use the generalized method of images
to find the Green's functions. Such a procedure was used to find the
Green's functions for the system with partially permeable wall
\cite{26} where the Green's function was obtained from
Eq.~(\ref{eq42a}) by reducing the IPS located at $-x_{0}$ by the
factor controlled by the permeability of the wall.

For the system with partially absorbing wall we start with a
physical condition, which can be stated as: \textit{if during a
given time interval $N$ particles reach the wall, the fraction
$\rho$ of them will be absorbed while $1-\rho$ will go through}.
The parameter $\rho$ is assumed to be a constant characterizing
the wall. Such a situation appears when the partially absorbing
wall is simulated by another IPS of the strength reduced by a
factor $\rho$. So, the Green's functions are as follows
    \begin{equation}\label{eq43}
 G_{A\,{\rm Dif}}(x,t;x_{0})=G_{0\,A}(x,t;x_{0})
-\rho_{A}G_{0\,A}(x,t;-x_{0}),
    \end{equation}
and
    \begin{equation}\label{eq44}
G_{B\,{\rm Dif}}(x,t;x_{0})=G_{0\,B}(x,t;x_{0})
-\rho_{B}G_{0\,B}(x,t;-x_{0}),
    \end{equation}
where
\begin{equation}\label{eq45}
G_{0\,i}(x,t;x_{0})=\frac{1}{\alpha|x-x_{0}|} H^{1 0}_{1 1}
   \left(\left(\frac{|x-x_{0}|}{\sqrt{D_{i}t^{\alpha}}}\right)
   ^{\frac{2}{\alpha}}\left|
           \begin{array}{cc}
             1 & 1 \\
             1 & 2/\alpha
           \end{array}\right. \right) ,
\end{equation}
for $i=A,B$. Using the integral formula (\ref{eq41}) and initial
conditions (\ref{eq9a}) and (\ref{eq9b}), we find (for details of the calculations
see the Appendix A)
    \begin{eqnarray}\label{eq46}
\lefteqn{C_{A\,{\rm Dif}}(x,t)=C_{0A}-\frac{2}{\alpha}\eta_{A}}\nonumber \\
&&\times H^{1 0}_{1 1}
\left(\left(\frac{-x}{\sqrt{D_{A}t^{\alpha}}}\right)^{2/\alpha}
            \left| \begin{array}{cc}
             1 & 1 \\
             0 & 2/\alpha
           \end{array}\right. \right),
\end{eqnarray}
where 
	\begin{equation}\label{eq46a}
\eta_{A}=C_{A0}(1+\rho_{A})/2 ,
	\end{equation}
and
    \begin{eqnarray}\label{eq47}
\lefteqn{C_{B\,{\rm Dif}}(x,t)=C_{0B}-\frac{2}{\alpha}\eta_{B}}\nonumber \\
&&\times H^{1 0}_{1 1}
\left(\left(\frac{x}{\sqrt{D_{B}t^{\alpha}}}\right)^{2/\alpha}
            \left| \begin{array}{cc}
             1 & 1 \\
             0 & 2/\alpha
           \end{array}\right. \right),
    \end{eqnarray}
where 
	\begin{equation}\label{eq47a}
\eta_{B}=C_{B0}(1+\rho_{B})/2 .
	\end{equation}
Let us note that when $C_{0A}=C_{0B}=C_0$ and $D_A=D_B$, we obtain $\rho_A=1$ and $\rho_B=1$ from Eqs.~(\ref{eq27}) and~(\ref{eq28}). The subdiffusive fluxes are given by the formula
    \begin{equation}\label{eq10}
J_{i}(x,t)=-D_{i}\frac{\partial^{1-\alpha}}{\partial
t^{1-\alpha}}\frac{\partial C_{i}(x,t)}{\partial x}.
    \end{equation}
Using Eqs.~(\ref{eq46}) and (\ref{eq47}), we obtain 
     \begin{eqnarray}\label{eq48}
\lefteqn{J_{A\,{\rm Dif}}(x,t)=\frac{2}{\alpha}\sqrt{D_{A}}\eta_{A}\left(\frac{\sqrt{D_{A}}}{-x}\right)^{\frac{2}{\alpha}-1}}\nonumber \\
&&\times H^{1 0}_{1 1}
  \left(\left(\frac{-x}{\sqrt{D_{A}t^{\alpha}}}\right)^{2/\alpha}
            \left| \begin{array}{cc}
             1 & 1 \\
             -1+2/\alpha & 2/\alpha
           \end{array}\right. \right),
    \end{eqnarray}
    \begin{eqnarray}\label{eq49}
\lefteqn{J_{B\,{\rm Dif}}(x,t)=-\frac{2}{\alpha}\sqrt{D_{B}}\eta_{B}\left(\frac{\sqrt{D_{B}}}{x}\right)^{\frac{2}{\alpha}-1}}\nonumber \\ &&\times H^{1 0}_{1 1}
  \left(\left(\frac{x}{\sqrt{D_{B}t^{\alpha}}}\right)^{2/\alpha}
            \left| \begin{array}{cc}
             1 & 1 \\
            -1+2/\alpha & 2/\alpha
           \end{array}\right. \right).
    \end{eqnarray}
In the following we use the shorter notation for the fluxes (\ref{eq48}) and (\ref{eq49})
	\begin{eqnarray}
\label{linja} J_{A \rm Dif}&=&\frac{\sqrt{D_A}\eta_A}{t^{1-\alpha/2}}Q\left(\frac{-x}{\sqrt{D_A t^{\alpha}}}\right) ,\\
\label{linjb} J_{B \rm Dif}&=&-\frac{\sqrt{D_B}\eta_B}{t^{1-\alpha/2}}Q\left(\frac{x}{\sqrt{D_B t^{\alpha}}}\right) ,
	\end{eqnarray}
where 
	\begin{equation}\label{defQ}
Q(z)=\frac{\alpha}{2}\sum^{\infty}_{k=0}\frac{1}{k!\Gamma(\alpha(1-k)/2)}(-z)^k.
	\end{equation}

\section{Time evolution of the reaction front}\label{terf}

In this section we derive the time evolution of the reaction front within the quasistatonary approximation. The derivation is based on three assumptions, which are expected to hold in the long time limit.
	\begin{enumerate}[(1)]
	\item\label{zalozenie1} We assume that the characteristic functions evolve in time according to the formulas
	\begin{equation}\label{zalWR}
W_R\sim t^{\alpha/6} ,
	\end{equation}	
\begin{equation}\label{zalDEP}
W_{{\rm Dep}}\sim t^{\alpha/2} ,
	\end{equation}	
	\begin{equation}\label{zalxf}
x_f(t)\sim t^{\alpha/2} .
	\end{equation}
The relations were derived in \cite{10} by means of the scaling method for the system where the subdiffusion coefficients of both reactants are equal to each other. The relation (\ref{zalWR}) was also found in \cite{ara} by means of the Monte Carlo simulations. Le us note that in Sec.~\ref{wrwd} we have shown that the relations (\ref{zalWR}) and (\ref{zalDEP}) are fulfilled for the system where $p=q=1$. The relation (\ref{zalxf}) will be also confirmed \textit{a posteriori} in this section.
	\item\label{zalozenie2} The region, where the quasistatic approximation works, extends beyond the reaction zone. Therefore there is the region defined by the relation
	\begin{equation}\label{defOVE}
W_R(t)\ll|x-x_{f}(t)|\ll W_{\rm Dep}(t) ,
	\end{equation}
where the quasistatic approximation region overlaps with the diffusion one.
	\item\label{zalozenie3} In the diffusion region the concentrations are given by Eqs. (\ref{eq46})--(\ref{eq47a}) with the parameters $\rho_A$ and $\rho_B$, which can be larger than unity.
	\end{enumerate}
Starting with the above assumptions, we show at first the following
	\begin{enumerate}[(a)]
	\item\label{z1}  The concentration profiles (\ref{eq46}) and (\ref{eq47}) extended to the reaction region vanish at the points which are identified with the point $x_z$ defined in Fig.~(\ref{Fig.1}) and by Eq.~(\ref{defxz}). In the long time limit the point $x_z$ is localized so close to $x_f$ that $x_z$ can be replaced by $x_f$.
	\item\label{z2} The fluxes $J_A$ and $J_B$ flowing into the reaction region from the left and from the right side, respectively, are assumed to be balanced in such a way that $m$ particles $A$ and $n$ particles $B$ flow into the reaction region in the time unit. 
\end{enumerate}
After showing that the conditions (\ref{z1}) and (\ref{z2}) hold, we use Eqs.~(\ref{eq46}) and (\ref{eq47}) to derive a relation describing the time evolution of the reaction front. 

As mentioned earlier, we are guided by the procedure already used for the normal diffusion--reaction systems \cite{12}. 
The quasistatic state can be defined by the following equations
    \begin{displaymath}
\frac{\partial^{1-\alpha}}{\partial t^{1-\alpha}}\left[D_{A}\frac{\partial^{2}}{\partial x^{2}}C_{A}(x,t)-mR(x,t)\right]=0,
    \end{displaymath}
and
    \begin{displaymath}
\frac{\partial^{1-\alpha}}{\partial t^{1-\alpha}}\left[D_{B}\frac{\partial^{2}}{\partial x^{2}}C_{B}(x,t)-nR(x,t)\right]=0,
    \end{displaymath}
which combined provide
    \begin{displaymath}
\frac{\partial^{1-\alpha}}{\partial t^{1-\alpha}}\frac{
\partial^{2}}{\partial x^{2}}\Psi(x,t)=0 ,
    \end{displaymath}
where    
	\begin{equation}\label{defPsi}
\Psi(x,t)\equiv\frac{1}{m}D_{A}C_{A}(x,t)-\frac{1}{n}D_{B}C_{B}(x,t) .
	\end{equation}
Using the formula (\ref{pochtni}), we find
    \begin{equation}\label{eq54}
\Psi(x,t)=E(x)t^{-\alpha}+F(t)x+G(t).
    \end{equation}
Applying the operator $\frac{\partial^{1-\alpha}}{\partial t^{1-\alpha}}\frac{\partial}{\partial x}$ to Eq.~(\ref{eq54}), we obtain
    \begin{equation}\label{eq55}
\frac{\partial^{1-\alpha}}{\partial t^{1-\alpha}}F(t)=\frac{1}{n}J_{B}(t)-\frac{1}{m}J_{A}(t).
    \end{equation}
The function $\Psi$ changes its sign in the
reaction zone from positive where $C_{B}\simeq 0$ to negative
where $C_{A}\simeq 0$. Thus, there is the point $x_{z}(t)$ which lies
inside the reaction zone, where the function $\Psi$ is equal to
zero. Therefore,
    \begin{equation}\label{defxz}
\Psi(x_{z}(t),t)=0.
    \end{equation}
Since $x_{f}(t)$ also lies inside the reaction zone there is
\begin{equation}\label{gwiazdka}
	|x_{z}(t)-x_{f}(t)|\leq\Omega t^{\alpha/6} ,
\end{equation}
where $\Omega$ is a positive constant. After simple calculations, we get
    \begin{equation}\label{eq57}
\Psi(x,t)=\frac{E(x)-E(x_{z}(t))}{t^{\alpha}}+F(t)(x-x_{z}(t)) .
    \end{equation}

Let us now consider the region where the region of diffusion approximation overlaps with the one of the quasistatic approximation for $x<x_f(t)$. The region occurs for such $x$ that the condition
	\begin{equation}\label{defOve}
-W_{\rm Dep}(x,t)\ll x-x_f(t) \ll -W_{\rm R}(x,t) ,
	\end{equation}
is fulfilled. Here $C_{A}\approx C_{A \rm Dif}$, $C_{B}\approx 0$, $J_{A}\approx J_{A \rm Dif}$, and $J_{B}\approx 0$. So, we get from Eq.~(\ref{defPsi})
	\begin{equation}\label{PsiOve}
\Psi(x,t)=\frac{1}{m}D_AC_{A \rm Dif}(x,t) ,
	\end{equation}
and from Eq.~(\ref{eq55}) 
	\begin{equation}\label{fjad}
\frac{\partial^{1-\alpha}}{\partial t^{1-\alpha}}F(t)=-\frac{1}{m}J_{A \rm Dif}(t) . 
	\end{equation} 
Let us note that $\Psi$ is given by the function of the variable $x/t^{\alpha/2}$ only (see Eq.~(\ref{eq46})). Therefore we deduce that
	\begin{equation}\label{defE}
E(x)=ax^2 ,
	\end{equation}
	\begin{equation}\label{defF}
F(t)=\frac{b}{t^{\alpha/2}} ,
	\end{equation}
and
	\begin{equation}\label{defG}
G(t)=c ,
	\end{equation}
where $a$, $b$, $c$ are constants.

We denote 
\begin{equation}\label{defx}
	x_f(t)-x=\epsilon(t).
\end{equation}
It is obvious that 
	\begin{displaymath}
\Omega_1 t^{\alpha/6}\ll \epsilon(t) \ll \Omega_2 t^{\alpha/2} ,
	\end{displaymath}
where $\Omega_1$ and $\Omega_2$ are positive constants.
When $t\rightarrow\infty$ the inequality provides $t^{\alpha/6}/\epsilon(t)\rightarrow 0$ and
	\begin{equation}\label{epsilon}
\epsilon(t)/t^{\alpha/2}\rightarrow 0 .
	\end{equation}
Combining Eqs. (\ref{eq46}), (\ref{eq57}), (\ref{PsiOve}) and  (\ref{defE})--(\ref{defx}) we obtain
\begin{eqnarray}\label{ala}
D_A\left[C_{0A}-\frac{2}{\alpha}\eta_A H^{1 0}_{1 1}
  \left(\left(\frac{\epsilon(t)-x_f(t)}{\sqrt{D_{A}t^{\alpha}}}\right)^{2/\alpha}
            \left| \begin{array}{cc}
             1 & 1 \\
             0 & 2/\alpha
           \end{array}\right. \right)\right]&=&\nonumber \\  
a\frac{(x_f(t)-\epsilon(t))^2-x_z^2(t)}{t^\alpha}
+b\frac{x_f(t)-\epsilon(t)-x_z(t)}{t^{\alpha/2}} .&&
\end{eqnarray}
Since in the long time limit $(x_f(t)-x_z(t)-\epsilon(t))/t^{\alpha/2}\rightarrow 0$ (see Eqs.~(\ref{gwiazdka}) and (\ref{epsilon})), from Eq.~(\ref{ala}) we get
	\begin{equation}\label{albb}
C_{0A}-\frac{2}{\alpha}\eta_A H^{1 0}_{1 1}
  \left(\left(\frac{-x_f(t)}{\sqrt{D_{A}t^{\alpha}}}\right)^{2/\alpha}
            \left| \begin{array}{cc}
             1 & 1 \\
             0 & 2/\alpha
           \end{array}\right. \right) = 0 .
	\end{equation}

Similar considerations performed in the region 
	\begin{displaymath}
W_R(x,t)\ll x-x_f(t)\ll W_{\rm Dep}(x,t)
	\end{displaymath}
provide
	\begin{displaymath}
\Psi(x,t)=-\frac{1}{n}D_BC_{B \rm Dif}(x,t) ,
	\end{displaymath}
and 
	\begin{equation}\label{aaaa}
\frac{\partial^{1-\alpha}}{\partial t^{1-\alpha}}F(t)=\frac{1}{n}J_{B \rm Dif}(t) ,
	\end{equation}
which gives
	\begin{equation}\label{abaa}
  C_{0B}-\frac{2}{\alpha}\eta_B H^{1 0}_{1 1}
  \left(\left(\frac{x_f(t)}{\sqrt{D_{B}t^{\alpha}}}\right)^{2/\alpha}
            \left| \begin{array}{cc}
             1 & 1 \\
             0 & 2/\alpha
           \end{array}\right. \right) = 0 .
	\end{equation}

From Eqs.~(\ref{fjad}) and (\ref{aaaa}) we obtain
	\begin{equation}\label{rowj}
\frac{1}{m}J_{A \rm Dif}=-\frac{1}{n}J_{B \rm Dif} ,
	\end{equation}
and from Eqs.~(\ref{linja}), (\ref{linjb}) and (\ref{rowj}) we get
\begin{equation}\label{rsl}
\frac{1}{m}\sqrt{D_A}\eta_A Q\left(\frac{-x_f(t)}{\sqrt{D_A t^{\alpha}}}\right)=\frac{1}{n}\sqrt{D_B}\eta_B Q\left(\frac{x_f(t)}{\sqrt{D_B t^{\alpha}}}\right).
\end{equation}
Combining Eqs. (\ref{albb}), (\ref{abaa}), (\ref{rsl}) and using the identity \cite{27}
	\begin{equation}\label{przekH}
H^{1 0}_{1 1}
  \left(z^{2/\alpha}\left| \begin{array}{cc}
             1 & 1 \\
             0 & 2/\alpha
           \end{array}\right. \right)=\frac{\alpha}{2}
           H^{1 0}_{1 1}
  \left(z\left| \begin{array}{cc}
             1 & \alpha/2 \\
             0 & 1
           \end{array}\right. \right) ,
	\end{equation}           
we have 
\begin{equation}\label{rowfi}
	\Phi\left(\frac{-x_f(t)}{\sqrt{D_A t^{\alpha}}}\right)=\frac{n}{m}\frac{\sqrt{D_A}C_{0A}}{\sqrt{D_B}C_{0B}} \Phi\left(\frac{x_f(t)}{\sqrt{D_B t^{\alpha}}}\right) ,
\end{equation}
where $\Phi(z)\equiv
H^{1 0}_{1 1}
  \left(z\left| \begin{array}{cc}
             1 & \alpha/2 \\
             0 & 1
           \end{array}\right. \right)/{Q(z)}$.
It is clear that there is only one point $x_f$ which for given $t$ fulfills the definition of reaction front. The solution of the Eq. (\ref{rowfi}) is
    \begin{equation}\label{xkt}
x_{f}(t)=Kt^{\alpha/2} ,
    \end{equation}
where coefficient $K$ is the solution of the following equation
	\begin{equation}\label{rowk}
	\Phi\left(\frac{-K}{\sqrt{D_A}}\right)=\frac{n}{m}\frac{\sqrt{D_A}C_{0A}}{\sqrt{D_B}C_{0B}} \Phi\left(\frac{K}{\sqrt{D_B}}\right) .
\end{equation}    
Thus, the time evolution of the reaction front is the power
function with the exponent depending on the subdiffusion parameter $\alpha$ only; the subdiffusion coefficients $D_A$ and $D_B$ controll the parameter $K$. Eqs. (\ref{xkt}) and (\ref{rowk}) are the main result of our paper. 

The procedure developed in this paper is a extension of the one
already used for the normal diffusion case \cite{12}. Repeating our consideration for $\alpha=1$ we obtain the results identical with those from \cite{12}. Our formula (\ref{xkt}) with $K$ given by Eq.~(\ref{rowk})
is a generalization of Eq.~(21) in \cite{10}.

\section{Numerical solutions}

To verify correctness of our procedure, we compare the analytical functions which are derived in the previous sections with numerical solutions. We show that there exists the quasistatic approximation zone where, as required by Eqs.~(\ref{eq54}) and (\ref{defE})--(\ref{defG}), the function $\Psi$ is parabolic with respect to $x$. We also show that there exists the region of overlap of the diffusion zone and the quasistatic one; in this region $C_{A \rm Dif}$ or $C_{B \rm Dif}$ are the linear functions of $x$. 

\subsection{Numerical procedure}\label{np}

As we show in the Appendix B, assuming that the functions $C_{A}$ and
$C_{B}$ and their second derivatives with respect to the space
variable are limited, Eq.~(\ref{eq2}) is equivalent to
    \begin{equation}\label{eq14}
\frac{^{C}\partial^{\alpha}}{\partial
t^{\alpha}}C_i(x,t)=D_{i}\frac{
\partial^{2}}{\partial x^{2}}C_i(x,t)-d_i R(x,t),
    \end{equation}
where $i=A,B$, $d_A=m$, $d_B=n$, with the Caputo fractional time derivative, which is defined
for $0<\alpha<1$ as \cite{20a}
	\begin{displaymath}
\frac{^{C}d^{\alpha}f(t)}{d
t^{\alpha}} =\frac{1}{\Gamma(1-\alpha)}\int_{0}^{t}d\tau
\frac{f(\tau)}{d \tau}(t-\tau)^{-\alpha}. 
	\end{displaymath}
Throughout this paper we denote the Riemann-Liouville fractional derivative without any additional index as $d^{\alpha}f(t)/dt^{\alpha}$, others kinds of the fractional derivatives are labeled by indexes $C$ for the Caputo fractional derivative and $GL$ for the Gr\"{u}nwald-Letnikov one.

In the papers \cite{gmmp,yuste} there were presented the procedures of the numerical solving of the subdiffusion equation without chemical reaction, when one can use the equation with Riemann-Liouville as well as Caputo fractional time derivative. The situation is different in the case of the subdiffusion-reaction equations. In the numerical calculations the fractional derivative is replaced by series. In the case of Eqs.~(\ref{eq2}) and (\ref{eq3}) with Riemann-Liouville fractional derivative we have relatively complicated expression under the derivative, whereas in Eq.~(\ref{eq14}) the fractional derivative acts only on concentrations. It caused that the numerical procedure based on Eq.~(\ref{eq14}) is more convenient to use, at least in our opinion.

To numerically solve the normal diffusion equation one usually
substitutes the time derivative by the backward difference
$\frac{\partial f(t)}{\partial t}\simeq \frac{f(t)-f(t-\Delta
t)}{\Delta t}$. In the presented procedure we proceed in a similar
way. We use the Gr\"{u}nwald-Letnikov fractional
derivative which is defined as a limit of a fractional-order
backward difference \cite{20a}
    \begin{equation}\label{gl}
\frac{^{GL}d^{\alpha}f(t)}{d
t^{\alpha}}=\lim_{\Delta t\rightarrow 0}(\Delta
t)^{-\alpha}\sum_{r=0}^{[\frac{t}{\Delta t}]}(-1)^{r} \left(
\begin{array}{c}
\alpha\\
r
\end{array} \right)f(t-r\Delta t),
    \end{equation}
where $\alpha>0$, $[z]$ means the integer part of $z$ and
\begin{eqnarray*}
\left( \begin{array}{c}
\alpha\\
r
\end{array} \right)&=&\frac{\Gamma(\alpha+1)}{r!\Gamma(\alpha-r+1)}\nonumber\\
&=&\frac{\alpha(\alpha-1)(\alpha-2)\cdot\ldots\cdot
[\alpha-(r-1)]}{1\cdot 2\cdot 3\cdot\ldots\cdot r}.
\end{eqnarray*}

When the function $f(t)$ of positive argument has continuous
derivatives of the first order, the
Riemann-Liouville fractional derivative is equivalent to the
Gr\"{u}nwald-Letnikov one for any parameter $\alpha$ ($0<\alpha<1$) \cite{20a}. So, we have
\begin{equation}\label{glrl}
\frac{d^{\alpha}f(t)}{d
t^{\alpha}}=\frac{^{GL}d^{\alpha}f(t)}{d
t^{\alpha}}.
\end{equation}
The relation between Riemann-Liouville and Caputo derivatives is
more complicated and reads as
\begin{equation}\label{rlc}
\frac{d^{\alpha}f(t)}{\partial
t^{\alpha}}=\frac{^{C}d^{\alpha}f(t)}{d
t^{\alpha}}+\Phi_{1-\alpha}(t)f(0),
\end{equation}
where
\begin{equation}\label{phi}
\Phi_{q+1}(t)=\left\{
\begin{array}{cc}
\frac{t^{q}}{\Gamma(q+1)}&t>0\\
0&t\leq 0
\end{array}
\right. .
\end{equation}
From Eqs. (\ref{gl})-(\ref{phi}) we can express the Caputo
fractional derivative in terms of the fractional-order backward
difference
\begin{eqnarray}\label{glc}
\frac{^{C}d^{\alpha}f(t)}{d t^{\alpha}}&=&\lim_{\Delta
t\rightarrow 0}(\Delta t)^{-\alpha}\sum_{r=0}^{[\frac{t}{\Delta
t}]}(-1)^{r}\left(
\begin{array}{c}
\alpha\\
r
\end{array}
\right)f(t-r\Delta t)\nonumber\\&&-\frac{1}{t^{\alpha}\Gamma(1-\alpha)}f(0).
\end{eqnarray}

The standard way to approximate of the fractional derivative,
which is useful for numerical calculations, is to omit the limit
in Eq.~(\ref{glc}) and to change the infinite series to the finite one
\begin{eqnarray}\label{glca}
\frac{^{C}d^{\alpha}f(t)}{d t^{\alpha}}&\simeq&(\Delta
t)^{-\alpha}\sum_{r=0}^{L}(-1)^{r} \left(
\begin{array}{c}
\alpha\\
r
\end{array}
\right)f(t-r\Delta t)\nonumber \\ &&-\frac{1}{t^{\alpha}\Gamma(1-\alpha)}f(0),
\end{eqnarray}
where the memory length $L$ is a natural number of arbitrary chosen value less then (or equal) $[t/\Delta t]$.

Subdiffusion is a process with the memory. According to the short memory principle, the fractional derivative is approximated by the fractional derivative with moving lower limit $t-L$, where $L$ is the 'memory length' equals to a certain amount of time steps \cite{20a}. However, in the paper \cite{lewkoszt} there was shown that the numerical solutions of subdiffusion equation with the boundary conditions (\ref{eq9a}) and (\ref{eq9b}) are in agreement with the analytical one only when the memory length is closed to actual account of time steps contrary to `short memory principle'. So, in numerical calculations we take the memory length $L$ equals to the actual number of time steps $t_s$.

Substituting Eq.~(\ref{glca}) to Eq.~(\ref{eq14}), using the
following approximation of the second order derivative 
\begin{displaymath}
\frac{d^{2}f(x)}{d x^{2}}\simeq \frac{f(x+\Delta
x)-2f(x)+f(x-\Delta x)}{(\Delta x)^{2}},
\end{displaymath}
we obtain
\begin{widetext}
\begin{eqnarray}\label{alg}
C_i(x,t)&=&-\sum_{r=1}^{L}(-1)^{r}\frac{\alpha(\alpha-1)(\alpha-2)
\cdot\ldots\cdot[\alpha-(r-1)]}{1\cdot 2\cdot 3\cdot\ldots\cdot r
}C_i(x,t-r\Delta t)+\frac{(\Delta t)^{\alpha}}{t^{\alpha}\Gamma(1-\alpha)}C_i(x,0)\nonumber \\
&& +D_{i}\frac{(\Delta
t)^{\alpha}}{(\Delta x)^{2}}[C_i(x+\Delta x,t-\Delta
t)-2C_i(x,t-\Delta t)+C_i(x-\Delta x,t-\Delta t)]\\
&&-d_i k(\Delta t)^{\alpha}C_A^m(x,t-\Delta t)C_B^n(x,t-\Delta t) \nonumber,
\end{eqnarray}
\end{widetext}
for $i=A,B$, $d_A=m$ and $d_B=n$.

\subsection{Numerical results}\label{nr}

Here we compare the analytical results with the numerical ones. In all figures there are presented functions calculated for the system where $\alpha=0.5$, $D_A=0.025$, $D_B=0.0125$, $C_{0A}=2$, $C_{0B}=1$, $k=1$, $m=n=1$. For numerical calculations we take $\Delta x=0.2$ and $\Delta t=0.05$ (all quantities are given in the arbitrary units). Additionally, in Figs.~(\ref{Fig.5}) and (\ref{Fig.6}) we plot the borders of the reaction zone $(x_f-W_R/2, x_f+W_R/2)$ calculated for the time $5000$. The position of the reaction front was calculated from the discrete version of Eq.~(\ref{defxf})
	\begin{equation}\label{xf}
x_f(t)=\frac{\sum_i x_i R(x_i,t)}{\sum_i R(x_i,t)} ,
	\end{equation}
and equals to $0.71$ for $t=5000$. The width of the reaction region calculated from discrete version of  Eq.~(\ref{defWR}) 
	\begin{displaymath}
W_{\rm R}^2(t)=\frac{\sum_i(x_i-x_f(t))^2 R(x_i,t)}{\sum_i R(x_i,t)} ,
	\end{displaymath}
equals to $0.38$ for $t=5000$. Thus the reaction region occupies the interval $(0.52;0.90)$. 

From Eq. (\ref{xf}) we find that
	\begin{equation}\label{1}
x_f(t)=0.0838t^{0.251}
	\end{equation}
This relation is very close to the relation (\ref{xkt}) with $K$ calculated from Eq.~(\ref{rowk}) which reads 
	\begin{equation}\label{2}
x_f(t)=0.0825t^{0.25}.  
	\end{equation}

\begin{figure}[h!]
\centering
\includegraphics[scale=0.84]{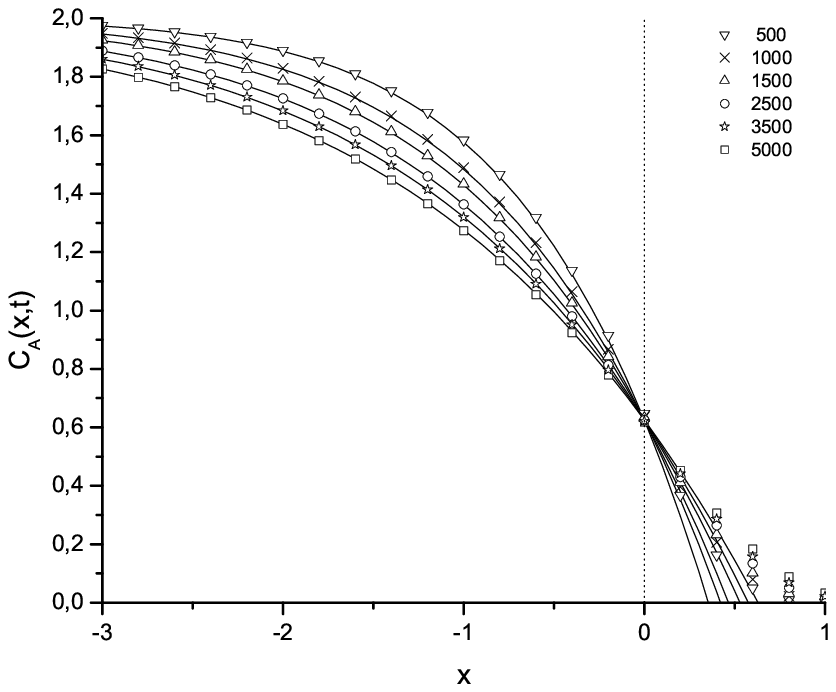}
\caption{\label{Fig.3}The symbols represent the numerical solutions of subdiffusion-reaction equation, the continuous lines are assigned to theoretical functions $C_{A {\rm Dif}}$ for the times given in the legend.}
\end{figure}

\begin{figure}[h!]
\centering
\includegraphics[scale=0.84]{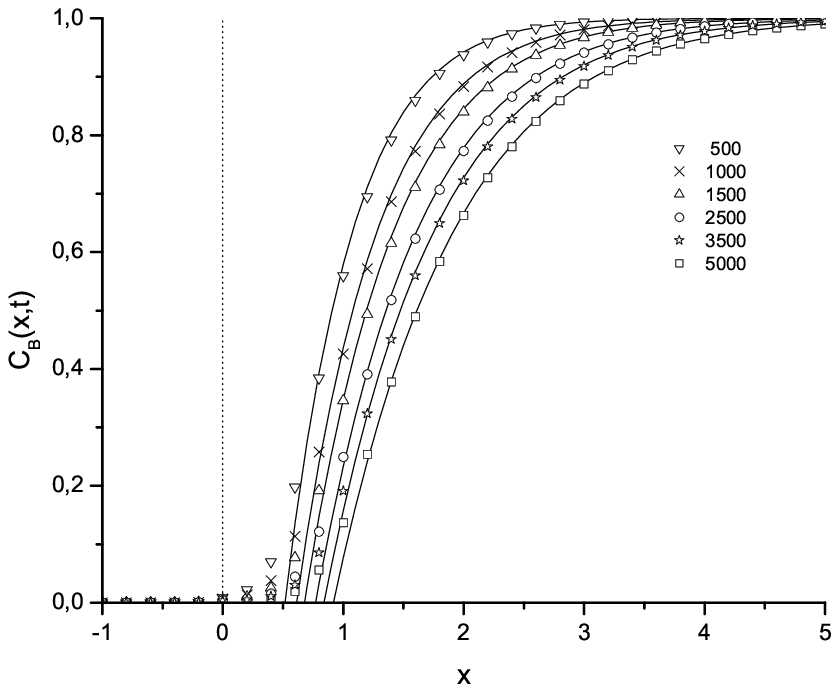}
\caption{\label{Fig.4}The symbols represent the numerical solutions of subdiffusion-reaction equation, the continuous lines are assigned to theoretical functions $C_{B {\rm Dif}}$ for the times given in the legend.}
\end{figure}

\begin{figure}[h!]
\centering
\includegraphics[scale=0.84]{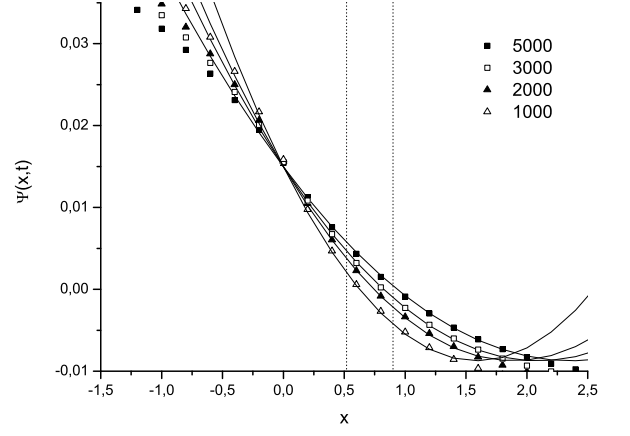}
\caption{\label{Fig.5}The function $\Psi$ (symbols) obtained numerically for the times given in the legend and their parabolic approximations inside the quasistatic approximation region (continuous line); the vertical lines represent the borders of the reaction zone calculated for $t=5000$.}
\end{figure}

\begin{figure}[h!]
\centering
\includegraphics[scale=0.82]{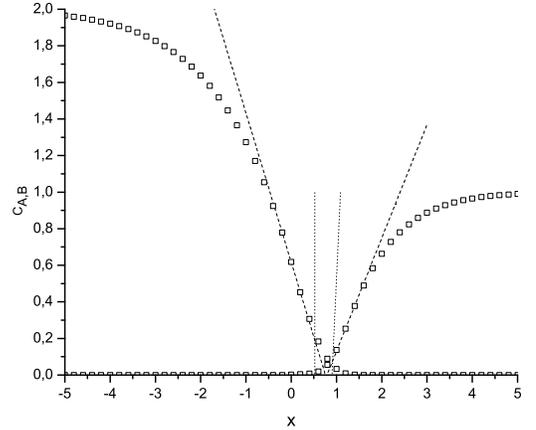}
\caption{\label{Fig.6}The concentration profiles $C_A$ and $C_B$ obtained numerically (squares) calculated for time $5000$ and their linear approximations (dashed lines), the vertical lines represent the borders of the reaction zone.}
\end{figure}

In Figs.~\ref{Fig.3} and \ref{Fig.4} there are presented the concentration profiles $C_A$ and $C_B$ obtained numerically according to the formula~(\ref{alg}) and the functions given by Eqs.~(\ref{eq46}) and (\ref{eq47}) with $\rho_A=0.40$ and $\rho_B=3.64$, respectively. We observe a quite good agreement of the analytical and numerical functions in the diffusion region.

In Fig.~\ref{Fig.5} we present the function $\Psi(x,t)$ calculated numerically and its parabolic approximation $\Psi(x,t)=0.297(x/t^{\alpha/2})^2-0.168(x/t^{\alpha/2})+0.015$. We note that $\Psi$ is satisfactorily approximated by the parabolic function of $x$. The region where $\Psi$ is parabolic determines the quasistatic approximation region. 

In Fig.~\ref{Fig.6} we present the numerical solutions of the subdiffusion--reaction equations and their linear approximations calculated from the formulas $C_A(x,t)\approx-0.816x+0.616$ and $C_B(x,t)\approx 0.620x-0.490$, respectively. As seen, the linear approximation of $C_A$ and $C_B$ is satisfactory outside the reaction region. This statement confirms correctness the quasistationary approximation in the region enclosing the reaction region. 

We conclude this section by saying that our numerical results support the postulates of the quasistatic approximation.

\section{Final remarks}\label{fr}

Using the quasistationary approximation and utilizing the solution of the subdiffusion--reaction equations in the diffusive region, we show that the time evolution of the reaction front for the subdiffusion--reaction
system is a power function (\ref{xkt}) with the exponent $\alpha/2$ and the coefficient $K$ is controlled by the subdiffusion coefficients of the system. 
The function $x_f\sim t^{\alpha/2}$ can be obtained by means of the scaling method. However, it is very hard within this method to find an explicit expression of the parameter $K$ for the case of $D_A\neq D_B$.

We note that in this paper we consider the process of subdiffusion controlls chemical reactions. It means that the reactions which proceed relatively fast when compared to the characteristic time of meeting of the particles of $A$ and $B$ \cite{11}. Under such assumption the quasistatic approximation works and the time evolution of the reaction front does not depend on the detailed form of the reaction term (expect of dependence of the parameters $m$ and $n$). This happens because the form of $R$ does not change the relation $W_R\sim t^{\alpha/6}$. Thus, the width of the reaction zone appears to be relatively small in comparison with the width of the quasistatic approximation region. The time evolution of $x_f$ is determined by the dynamics of transport of the particles to the reaction zone and it depends on the parameters $m$, $n$, $D_A$, $D_B$, $C_{0A}$ and $C_{0B}$ only. This statement is particularly important for the subdiffusion--reaction systems where the reaction term is not uniquely defined (as the fractional derivative can be involved into this term in a few ways \cite{27a,10,9,sbsl,s,hw,10a}). 

As far as we know, the time evolution of the reaction front has not been measured experimentally in a subdiffusive system with two mobile reactants. For this reason we can compare the functions (\ref{eq46}) and (\ref{eq47}) with experimental data obtained for a subdiffusive system without chemical reactions. Our theoretical and the experimental functions presented in \cite{dsdoww} are qualitatively similar to each other if we take the units commonly used in real systems where $x$ is given in $10^{-2} m$, $t$ in $sec$, $D_A$ and $D_B$ are of the order $10^{-8} m^2/s^\alpha$. Since $K$ is controlled by the subdiffusion coefficients of reactants, the method presented in this paper can be used for extracting the subdiffusion parameter from experimental data. The numerical calculations show that if we take the subdiffusion coefficients of the order maintained above and we assume that $C_{0A}/C_{0B}$ is of the order of $1$, we obtain $K\sim 10^{-2}$ $m/s^{\alpha/2}$ from Eq.~(\ref{rowk}).

The quasistatic approximation in a normal diffusion system applies to a region where the equilibrium time $\tau_{\rm F}$ of the reaction region is negligibly small comparing to the characteristic time of change of the flux $\tau_{\rm J}$ in the long time limit \cite{12,ckd,cd}. Let us note that this fact is fulfilled in the subdiffusive--recation system.
Since $W_{\rm R}\sim t^{\alpha/6}$, we have $\tau_{\rm F}\sim t^{1/3}$ form (\ref{tf}). Taking the definition (\ref{tj}), which for the subdiffusion flux $J\sim 1/t^{1-\alpha/2}$ gives $\tau_{\rm J}\sim 1/t$ (see Eqs. (\ref{linja}) and (\ref{linjb})), we get $\tau_{\rm F}/\tau_{\rm J}\rightarrow_{t\rightarrow\infty}0$ for any value of the subdiffusive parameter $\alpha$. So, the assumptions adopted in our paper agree with the quasistationary condition (\ref{r4}). 

The function $\Psi$ is approximated by parabolic function in the region where the quasistatic approximation region overlaps with the diffusion one. However, we expect that there are departures from this approximation in a region located within the reaction zone where the reaction term is significantly different from zero. It is because of the concentrations $C_A$ and $C_B$ have different scaling properties in that region. We expect that the width of that region is so narrow, as compared with the width the quasistatic approximation one, that the departure form parabolic approximation is hard to observe on the plots presented in our paper. We note that the possibility of occurring this departure does not influence our main results.

\begin{acknowledgments}
The authors wish to express their thanks
to Stanis{\l}aw Mr\'{o}wczy\'{n}ski for fruitful discussion and
critical comments on the manuscript. This paper was supported by
Polish Ministry of Education and Science under Grant No. 1 P03B 136
30.
\end{acknowledgments}

\appendix

\section{}

In this Appendix we present some details of the procedure of
solving the subdiffusion equation. The calculations with
Riemann--Liouville fractional time derivative are relatively simple in
terms of the Laplace transform (LT) $\hat{L}[f(t)]\equiv
\hat{f}(s)=\int^{\infty}_{0}dtf(t)e^{-st}$. The LT of Green's
function for a homogeneous subdiffusive system without chemical
reaction (\ref{eq45}) reads \cite{23,mk}
\begin{equation}\label{a1}
\hat{G}_{0,i}(x,s;x_{0})=\frac{1}{2\sqrt{D_{i}}s^{1-\alpha/2}} e^{-|x-x_{0}|\sqrt{s^{\alpha}/D_{i}}} ,
\end{equation}
$i=A,B$. The LT commutes with the integration (whith respect to the
variable $x$), so from Eq.~(\ref{eq41}) we get
    \begin{equation}\label{a2}
\hat{C}(x,s)=\int \hat{G}(x,s;x_{0})C(x_{0},0)dx_{0} .
    \end{equation}
Putting (\ref{a1}) to the LT of Eqs.~(\ref{eq43}) and (\ref{eq44}),
and next to Eq.~(\ref{a2}), we obtain
    \begin{equation}\label{a3}
\hat{C_{A}}(x,s)=\frac{C_{0A}}{s}-\frac{\eta_{A}}{s}e^{-(-xs^{\alpha/2})/\sqrt{D_{A}}} ,
    \end{equation}
with $\eta_{A}=C_{0A}(1+\rho_{A})/2$, and
    \begin{equation}\label{a4}
\hat{C_{B}}(x,s)=\frac{C_{0B}}{s}-\frac{\eta_{B}}{s}e^{-xs^{\alpha/2}/\sqrt{D_{B}}} ,
    \end{equation}
with $\eta_{B}=C_{0B}(1+\rho_{B})/2$. The inverse Laplace transform
$\hat{L}^{-1}$ gives (here $a>0$ and $\beta>0$) \cite{23}
    \begin{equation}\label{a5}
\hat{L}^{-1}\left( s^{\nu}e^{-as^{\beta}}\right)=\frac{1}{\beta
a^{(1+\nu)/\beta}}H^{1 0}_{1 1}
  \left(\frac{a^{1/\beta}}{t}
            \left| \begin{array}{cc}
             1 & 1 \\
             \frac{1+\nu}{\beta} & \frac{1}{\beta}
           \end{array}\right. \right) ,
    \end{equation}
and
    \begin{equation}\label{a6}
\hat{L}^{-1}\left(s^{\nu}e^{-as^{\beta}}\right)=\frac{1}{t^{1+\nu}}\sum_{j=0}^{\infty}
\frac{1}{j!\Gamma(-\nu-j\beta)}\left(-\frac{a}{t^{\beta}}\right)^{j}.
    \end{equation}
Using the relation (\ref{a5}) to calculate the inverse LT of
Eqs.~(\ref{a3}) and (\ref{a4}) we get (\ref{eq46}) and (\ref{eq47}).
Let us note that comparing the right hand sides of Eqs.~(\ref{a5})
and (\ref{a6}), after simple calculations we get the useful relation
(\ref{eq29}).

The LT of the subdiffusive flux (\ref{eq10}) reads
    \begin{equation}\label{a7}
\hat{J}_{i}(x,s)=-D_{i}s^{1-\alpha}\frac{d\hat{C_i}(x,s)}{dx}
.
    \end{equation}
Putting Eqs.~(\ref{a3}) and (\ref{a4}) to (\ref{a7}) and next to
Eq.~(\ref{a5}), we obtain (\ref{eq48}) and (\ref{eq49}).

\section{}

Here we show that Eqs. (\ref{eq2}) and (\ref{eq14}) are equivalent
to each other when the concentration $C$ and its second space derivative are limited.
The Laplace transforms of
fractional derivatives are as follows \cite{20a} (here
$0<\alpha<1$)
    \begin{displaymath}
\hat{L}\left[\frac{d^{\alpha}f(t)}{dt^{\alpha}}\right]=\left.
s^{\alpha}\hat{f}(s)-
\frac{d^{\alpha-1}f(t)}{dt^{\alpha-1}}\right|_{t=0},
    \end{displaymath}
    \begin{displaymath}
\hat{L}\left[\frac{^{C}d^{\alpha}f(t)}{dt^{\alpha}}\right]=
s^{\alpha}\hat{f}(s)-s^{\alpha-1}f(0).
    \end{displaymath}
The Laplace transform of Eq. (\ref{eq2}) is
    \begin{eqnarray}\label{b3}
\lefteqn{s\hat{C}(x,s)-C(x,0)=}\nonumber \\
&&s^{1-\alpha}\left[D\frac{d^{2}\hat{C}(x,s)}{dx^{2}}
-\hat{R}(x,s)\right] \\
&&-\frac{d^{-\alpha}}{dt^{-\alpha}}\left.\left[D
\frac{d^{2}C(x,t)}{dx^{2}}-R(x,t)\right]\right|_{t=0},\nonumber
    \end{eqnarray}
whereas the Laplace transform of Eq. (\ref{eq14}) reads as
$s^{\alpha}\hat{C}(x,s)-s^{\alpha-1}C(x,0)=D\frac{d^{2}\hat{C}(x,s)}{dx^{2}}
-\hat{R}(x,s)$, which gives
    \begin{equation}\label{b4}
s\hat{C}(x,s)-C(x,0)=s^{1-\alpha}\left[D\frac{d^{2}\hat{C}(x,s)}{dx^{2}}
-\hat{R}(x,s)\right].
    \end{equation}
We assume that the function $C$ and its second space-variable
derivative are limited. So, there is a positive number $M$ which
fulfils the relation $|\Theta (x,t)|<M$, where $\Theta (x,t)\equiv
D \frac{d^{2}C(x,t)}{dx^{2}}-R(x,t)$, for any $(x,t)$. 
From the definition of Riemann-Liouville derivative of negative
order
\begin{displaymath}
\frac{d^{-\alpha}}{dt^{-\alpha}}f(t)=\frac{1}{\Gamma(\alpha)}\int_{0}^{t}d\tau
(t-\tau)^{\alpha-1}f(\tau),
\end{displaymath}
we obtain
\begin{equation}\label{b6}
\left|\frac{d^{-\alpha}}{dt^{-\alpha}}\Theta(x,t)\right|\leq
M\int_{0}^{t}d\tau
(t-\tau)^{\alpha-1}=\frac{M}{\alpha}t^{\alpha}.
    \end{equation}
and from Eq. (\ref{b6}) we get
    \begin{displaymath}
\frac{d^{-\alpha}}{dt^{-\alpha}}\left.\Theta(x,t)\right|_{t=0}=0 ,
    \end{displaymath}
which causes that the Laplace transforms (\ref{b3}) and (\ref{b4})
are equal to each other. Thus, the equations (\ref{eq2}) and
(\ref{eq14}) are equivalent to each other for the limited function $\Theta$.

\end{document}